\documentclass[reprint,prl,twocolumn,showpacs,superscriptaddress,amssymb,amsmath,preprintnumbers,floatfix]{revtex4-1}

\usepackage{amsmath}
\usepackage{amssymb}
\usepackage{bm}
\usepackage{dcolumn}
\usepackage{graphicx}
\usepackage{rotating}
\usepackage{subfigure}
\usepackage{natbib}

\begin{document}

\date{\today}
\preprint{RIKEN-QHP-25}
\title{%
Tenth-Order QED Contribution to the Electron {\boldmath $g\!-\!2$} \\
and an Improved Value of the Fine Structure Constant
}

\author{Tatsumi Aoyama}
\affiliation{Kobayashi-Maskawa Institute for the Origin of Particles and the Universe (KMI), Nagoya University, Nagoya, 464-8602, Japan}
\affiliation{Nishina Center, RIKEN, Wako, Japan 351-0198 }

\author{Masashi Hayakawa}
\affiliation{Department of Physics, Nagoya University, Nagoya, Japan 464-8602 }
\affiliation{Nishina Center, RIKEN, Wako, Japan 351-0198 }

\author{Toichiro Kinoshita}
\affiliation{Laboratory for Elementary Particle Physics, Cornell University, Ithaca, New York, 14853, U.S.A }
\affiliation{Nishina Center, RIKEN, Wako, Japan 351-0198 }

\author{Makiko Nio}
\affiliation{Nishina Center, RIKEN, Wako, Japan 351-0198 }


\begin{abstract}

This paper presents the complete QED contribution to the electron $g\!-\!2$ up to the tenth order. 
With the help of the automatic code generator, 
we have evaluated all 12672 diagrams of the tenth-order diagrams 
and obtained $9.16 ~(58)(\alpha/\pi)^5$. 
We have also improved the eighth-order contribution obtaining 
$-1.9097~(20)(\alpha/\pi)^4$, 
which includes the mass-dependent contributions.
These results lead to  $a_e(\text{theory}) 
=1~159~652~181.78~(77)\times 10^{-12}$.
The improved value of the fine-structure constant $\alpha^{-1} 
= 137.035~999~174~(35)~[0.25 \text{ppb}]$ is also derived 
from the theory and measurement of $a_e$. 


\end{abstract}

%

\maketitle


The anomalous magnetic moment $a_e \equiv (g\!-\!2)/2$ of the electron has played 
the central role in testing the validity of quantum electrodynamics (QED)
as well as the standard model of the elementary particles.  
On the experimental side
the measurement of $a_e$ by the Harvard group has reached 
the astonishing precision
\cite{Hanneke:2008tm,Hanneke:2010au}:
\begin{eqnarray}
a_e(\text{HV})= 1~159~652~180.73~ (0.28) \times 10^{-12}~[0.24 \text{ppb}]
~.
\label{a_eHV08}
\end{eqnarray}

In the standard model the contribution to $a_e$ comes from three 
types of interactions, electromagnetic, hadronic, and electroweak:
\begin{equation}
a_e = a_e (\text{QED}) + a_e (\text{hadronic}) + a_e (\text{electroweak}).
\label{electronanomaly}
\end{equation}
The QED contribution can be evaluated by the perturbative expansion
in $\alpha/\pi$:
\begin{equation}
a_e ({\rm QED}) = \sum_{n=1}^\infty \left( \frac{\alpha}{\pi} \right)^{n} a_e^{(2n)},
\label{aeQED}
\end{equation}
where $a_e^{(2n)}$ is finite due to the renormalizability of QED 
and may be written  
in general as
\begin{align}
a_e^{(2n)}& = A_1^{(2n)}+A_2^{(2n)} (m_e/m_\mu)+A_2^{(2n)} (m_e/m_\tau)
\nonumber \\
&+A_3^{(2n)} (m_e/m_\mu,m_e/m_\tau) 
\label{QEDterm}
\end{align}
to show the mass-dependence explicitly.
We use the latest values of the electron-muon mass ratio 
$m_e/m_\mu = 4.836~331~66~(12) \times 10^{-3}$ 
and the electron-tau mass ratio $m_e/m_\tau = 2.875~92~(26) \times 10^{-4}$
\cite{Mohr:2012tt}. 

\begin{figure}[t]
\includegraphics[width=8.4cm]{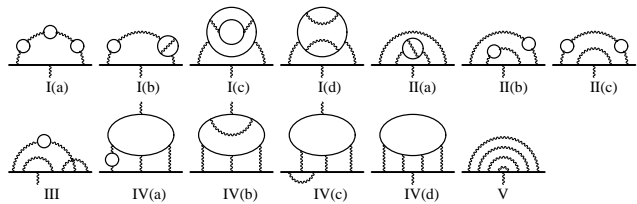}
\caption{
\label{fig:8th}
Typical vertex diagrams representing 13
gauge-invariant subsets contributing to the eighth-order lepton $g\!-\!2$.
}
\vspace{-0.43cm}
\end{figure}

The first three terms of $A_1^{(2n)}$ are known analytically 
\cite{Schwinger:1948iu,Petermann:1957,Sommerfield:1958,Laporta:1996mq}, while
$A_1^{(8)}$ and $A_1^{(10)}$ are known only by numerical integration 
\cite{Kinoshita:2005zr,Aoyama:2007dv,*Aoyama:2007mn}.
They are summarized as:
\begin{align}
& A_1^{(2)} = 0.5,  \nonumber \\
& A_1^{(4)} = -0.328~478~965~579~193~\ldots,  \nonumber \\
& A_1^{(6)} =  1.181~241~456~\ldots,  \nonumber \\
& A_1^{(8)} = -1.9106~(20),~  
\label{Eq:A_1(8)}
\\ 
& A_1^{(10)} = 9.16~(58)~. 
\label{Eq:A_1(10)}
\end{align}
The $A_1^{(8)}$ is obtained from 891 Feynman diagrams 
classified into 13 gauge-invariant subsets (see Fig.~\ref{fig:8th}). 
The value $A_1^{(8)}=-1.9144~(35)$ in \cite{Aoyama:2007mn}  
was confirmed by the new calculation and replaced by the
updated value (\ref{Eq:A_1(8)}). 

\begin{figure}[t]
\includegraphics[width=8.2cm]{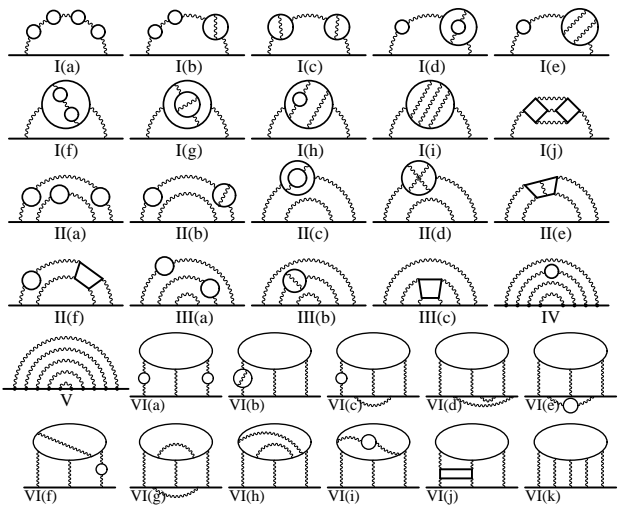}
\caption{
\label{fig:10th}
Typical self-energy-like diagrams representing 32
gauge-invariant subsets contributing to the tenth-order lepton $g\!-\!2$.
Solid lines represent lepton lines propagating in a 
weak magnetic field. }
\vspace{-0.43cm}
\end{figure}

The $A_1^{(10)}$ receives the contribution from 12672 diagrams classified into
 32 gauge-invariant subsets (see Fig.~\ref{fig:10th}).  
The results of 31 gauge-invariant subsets have been 
published\cite{Kinoshita:2005sm,Aoyama:2008gy, Aoyama:2008hz,Aoyama:2010yt, Aoyama:2010pk,Aoyama:2011rm,Aoyama:2010zp,Aoyama:2011zy,
Aoyama:2011dy,Aoyama:2012fc}.
The remaining set, Set V,  
consists of 6354 diagrams, which are 
more than half of all tenth-order diagrams.
However, we have managed to evaluate it 
\cite{Aoyama:2012setV}
with a precision which leads to theory more accurate than that of the measurement
(\ref{a_eHV08}):
\begin{equation}
  A_1^{(10)}[\text{Set V}] = 10.092~ (570) .
\label{Eq:A_1(10)SetV}
\end{equation}
Adding data of all  32 gauge-invariant subsets, we are now able 
to obtain the complete value of $A_1^{(10)}$ as in (\ref{Eq:A_1(10)}),
which replaces the crude estimate $A_1^{(10)}=0.0(4.6)$ 
\cite{CODATA:1998,Gabrielse:2006gg,*Gabrielse:2006ggE,Mohr:2012tt}. 

The mass-dependent terms $A_2$ and $A_3$  of the fourth and sixth orders
are known \cite{Elend:1966a,*Elend:1966b,Samuel:1990qf,Li:1992xf,Laporta:1992pa,Laporta:1993ju, Passera:2006gc}
and re-evaluated using the updated mass ratios \cite{Mohr:2012tt},
\begin{align}
& A_2^{(4)}(m_e/m_\mu)   = 5.197~386~67~(26)~\times 10^{-7},  \nonumber \\
& A_2^{(4)}(m_e/m_\tau)  = 1.837~98~(34)~\times 10^{-9},  \nonumber \\
& A_2^{(6)}(m_e/m_\mu)   = -7.373~941~55~(27)~\times 10^{-6},  \nonumber \\
& A_2^{(6)}(m_e/m_\tau)  = -6.583~0~(11)~\times 10^{-8}, \nonumber \\
& A_3^{(6)}(m_e/m_\mu,m_e/m_\tau) =  0.1909~(1)~\times 10^{-12} .
\end{align}
Except for $A_3^{(6)}$ all are known analytically so that
the uncertainties come only from fermion-mass ratios.

The mass-dependent terms of the eighth order
and the muon contribution to the tenth order are numerically evaluated 
\cite{Aoyama:2008gy, Aoyama:2008hz,Aoyama:2010yt, Aoyama:2010pk,Aoyama:2011rm,Aoyama:2010zp,Aoyama:2011zy,Aoyama:2011dy,Aoyama:2012fc}.
Our new results are summarized as 
\begin{align}
& A_2^{(8)}(m_e/m_\mu)   =  9.222~(66)~\times 10^{-4},  \nonumber \\
& A_2^{(8)}(m_e/m_\tau)  =  8.24~(12)~\times 10^{-6}, \nonumber \\
& A_3^{(8)}(m_e/m_\mu,m_e/m_\tau) =  7.465~(18)~\times 10^{-7} ,
\nonumber \\
& A_2^{(10)}(m_e/m_\mu)  = -0.003~82~(39).
\label{Eq:A_2(10)}
\end{align}

The hadronic contribution to $a_e$ is summarized in Ref.~\cite{Mohr:2012tt}. 
The leading order \cite{Davier:1998si} and 
next-to-leading order (NLO) \cite{Krause:1996rf} contributions 
of the hadronic vacuum-polarization (v.p.) 
as well as 
the hadronic light-by-light-scattering (\textit{l-l}) term \cite{Prades:2009tw} 
are given as
%
\begin{eqnarray}
a_e (\text{had.~v.p.})    &=& 1.875~(18) ~\times 10^{-12}, \nonumber \\
a_e (\text{NLO~had.~v.p.}) &=&-0.225~(5) ~\times 10^{-12}, \nonumber \\
a_e (\text{had.~{\it l-l}})     &=& 0.035~(10) ~\times 10^{-12}.
\label{Eq:hadronic}
\end{eqnarray}
At present no direct evaluation of the two-loop electroweak effect is available.
The best estimate is the one obtained by
scaling down from the electroweak effect on $a_\mu$
\cite{Fujikawa:1972fe,Czarnecki:1995sz,Knecht:2002hr,Czarnecki:2002nt}:
\begin{equation}
a_e ({\rm weak}) = 0.0297~(5) ~\times 10^{-12}.
\label{weak}
\end{equation}

To compare the theoretical prediction with the measurement 
(\ref{a_eHV08}),
we  need  the value of the fine-structure constant $\alpha$
determined by a method independent of $g\!-2\!$ .
The best $\alpha$ available at present is the one 
obtained from the measurement of $h/m_{\text{Rb}}$ \cite{Bouchendira:2010es}, 
combined with the very precisely known Rydberg constant 
and $m_\text{Rb}/m_e$ \cite{Mohr:2012tt} :
\begin{eqnarray}
\alpha^{-1} (\text{Rb10}) = 137.035~999~049~(90)~~~[0.66 \text{ppb}].
\label{alinvRb10}
\end{eqnarray}  
With this  $\alpha$ the theoretical prediction of $a_e$ becomes 
\begin{align}
a_e(\text{theory}) = & 1~159~652~181.78~(6)(4)(3)(77) 
\nonumber \\
                     & ~~~~~~~~ \times 10^{-12}~~~[ 0.67\text{ppb}],
\label{a_etheory}
\end{align}
where the first, second, third, and fourth uncertainties come
from the  eighth-order term (\ref{Eq:A_1(8)}),
the tenth-order  term (\ref{Eq:A_1(10)}),
the hadronic corrections (\ref{Eq:hadronic}), and the
fine-structure constant (\ref{alinvRb10}), respectively.
This is in good agreement with the
experiment (\ref{a_eHV08}):  
\begin{eqnarray}
a_e(\text{HV}) - a_e(\text{theory}) = -1.06~ (0.82) \times 10^{-12}.
\label{exp-theory}
\end{eqnarray}
More rigorous comparision between experiment and  theory is 
hindered by the uncertainty of $\alpha^{-1}(\text{Rb})$ in (\ref{alinvRb10}). 
Note that the sum $1.685 (21) \times 10^{-12}$ of 
the hadronic contributions (\ref{Eq:hadronic}) is 
now larger than Eq.~(\ref{exp-theory}).  
It is thus desirable to reexamine and update the values of 
the hadronic contributions.

The equation (\ref{a_etheory}) shows clearly that
the largest source of uncertainty
is the fine-structure constant (\ref{alinvRb10}).
To put it differently, it means that a non-QED $\alpha$, even the best one
available at present,
is too crude to test QED to the extent achieved by the theory
and measurement of $a_e$.
Thus it makes more sense to test QED by an alternative approach, namely,
compare $\alpha^{-1}$(Rb10) with
$\alpha^{-1}$ obtained from theory and measurement of $a_e$.
This leads to
\begin{equation}
\alpha^{-1}(a_e) = 137.035~999~1736~(68)(46)(26)(331)~~~[0.25 \text{ppb}],
\label{alinvae}
\end{equation}
where the first, second, third, and fourth uncertainties come
from the eighth-order and 
the tenth-order QED terms, 
the hadronic and electroweak terms, and the
measurement of $a_e$(HV) in (\ref{a_eHV08}), respectively.
The uncertainty due to theory  has been 
improved  by a factor 4.5 compared with 
the previous one \cite{Gabrielse:2006gg}.

\begingroup
\renewcommand{\baselinestretch}{1.17}
 \begin{table*}
 \caption{
 The eighth-order QED contribution from 13 gauge-invariant groups to
 electorn $g\!-\!2$. The values with a superscript $a$, $b$, or $c$  
are quoted from Refs.\cite{Caffo:1978mg}, \cite{Kinoshita:2005zr}, 
or \cite{Kinoshita:2002ns}, respectively. 
$n_f$ shows the number of vertex diagrams 
contributing to $A_1^{(8)}$.
Other values are obtained from evaluation of new programs.
The mass-dependence of $A_3^{(8)}$ is $A_3^{(8)}(m_e/m_\mu,m_e/m_\tau)$.
\label{Table:electron8thorder}
 }
 \begin{ruledtabular}
 \begin{tabular}{l@{\hskip-0em}r@{\hskip-5em}d
                  @{\hskip-5em}d@{\hskip-5em}d@{\hskip-5em}d}
   group &  $n_f$   
&\makebox[-6em]{$A_1^{(8)}$}
&\makebox[-7em]{$A_2^{(8)}(m_e/m_\mu)\times 10^3$}
&\makebox[-7em]{$A_2^{(8)}(m_e/m_\tau)\times 10^5$}
&\makebox[-3em]{$A_3^{(8)}\times 10^7$}
\\
 \hline
 I(a) & 1  &   0.000~876~865~\cdots ^a
           &   0.000~226~456~(14) 
           &   0.000~080~233~(5)    
           &   0.000~011~994~(1) \\
 I(b) & 6  &   0.015~325~20~(37)   
           &   0.001~704~139~(76) 
           &   0.000~602~805~(26)   
           &   0.000~014~097~(1) \\
 I(c) & 3  &   0.011~130~8~(9)^b    
           &   0.011~007~2~(15)  
           &   0.006~981~9~(12)     
           &   0.172~860~(21) \\
 I(d) & 15 &   0.049~514~8~(38)    
           &   0.002~472~5~(7)    
           &   0.087~44~(1)         
           &   0     \\
 II(a)& 36 &  -0.420~476~(11)      
           &  -0.086~446~(9)       
           &  -0.045~648~(7)      
           &   0       \\
 II(b)& 6  &  -0.027~674~89~(74)   
           &  -0.039~000~3~(27)   
           &  -0.030~393~7~(42)     
           &  -0.458~968~(17)  \\ 
 II(c)& 12 &  -0.073~445~8~(54)    
           &  -0.095~097~(24)     
           &  -0.071~697~(25)       
           &  -1.189~69~(67) \\
 III  &150 &   1.417~637~(67)      
           &   0.817~92~(95)      
           &   0.6061~(12)          
           &   0      \\  
 IV(a)& 18 &   0.598~838~(19)      
           &   0.635~83~(44)      
           &   0.451~17~(69)      
           &   8.941~(17) \\ 
 IV(b)& 60 &   0.822~36~(13)       
           &   0.041~05~(93)      
           &   0.014~31~(95)         
           &   0      \\
 IV(c)& 48 &  -1.138~52~(20)       
           &  -0.1897~(64)      
           &  -0.102~(11)            
           &   0      \\
 IV(d)& 18 &  -0.990~72~(10)^c       
           &  -0.1778~(12)        
           &  -0.0927~(13)                       
           &   0      \\
 V    &518 &  -2.1755~(20)         
           &   0                 
           &   0                   
           &   0      \\
 \end{tabular}
 \end{ruledtabular}
 \end{table*}
\renewcommand{\baselinestretch}{1.00}
\endgroup

Let us now discuss the eighth- and tenth-order
calculations in more details.
The 13 gauge-invariant groups of the eighth order
were numerically evaluated  by VEGAS \cite{Lepage:1977sw}
and published \cite{Kinoshita:2005sm,Aoyama:2007dv,*Aoyama:2007mn}.
As an independent check, we built all programs of the 12 groups 
from scratch with the help of automatic code generator {\sc gencode}{\it N},
except for Group IV(d) which had already been calculated  
by two different methods \cite{Kinoshita:2002ns}.
The new values of the mass-independent contributions of 
all 12 groups are consistent with the old values. We have 
thus statistically combined
two values and listed the results in Table~I. 
Since the validity of the new programs were
confirmed in this way, we used the new programs as well as the old programs
to evaluate the mass-dependent terms $A_2^{(8)}$ and $A_3^{(8)}$. 

Group V deserves a particular attention which
consists of 518 vertex diagrams and 
is the source of the largest uncertainty of $a_e^{(8)}$. 
The programs generated by {\sc gencode}{\it N}  have been
evaluated with intense numerical work which led to
$- 2.173~77~(235)$. This is
consistent with the value in \cite{Aoyama:2007dv, *Aoyama:2007mn},
$- 2.179~16~(343)$. The combined value is 
\begin{equation}
A_1^{(8)}[\text{Group V}] = - 2.175~50~(194)~ .
\end{equation}
This improvement results in about 40\% reduction of the uncertainty of the
eighth-order term.

The tenth-order contribution comes from 32 gauge-invariant subsets (see Fig.~\ref{fig:10th}).
The {\sc fortran} programs  
of integrals of 15 subsets I(a-f), II(a,b), II(f), VI(a-c), VI(e,f), and VI(i)
are straightforward
and obtained  by a slight modification of programs for the eighth-order diagrams. 
Together with the results of subsets VI(j,k), the contributions from 17 
subsets to $A_1^{(10)}$ were evaluated and published \cite{Kinoshita:2005sm}.
We recalculated all 17 subsets once more from scratch and 
found that the results of I(d), I(f), II(a), II(b), and VI(c) in
\cite{Kinoshita:2005sm} were incorrect. Although the constructed integrals 
for the first four subsets  are free from errors, 
they did not include the finite renormalization terms 
in the last step of the calculation.  The value of the subset VI(c) was a typo.
The corrected values  are  listed in Table~II.

Other subsets are far more difficult to handle.
Thus we developed and utilized the code-generating algorithm
{\sc gencode}{\it N} which 
carries out all steps automatically,
including subtraction of ultraviolet and infrared divergences
\cite{Aoyama:2005kf,*Aoyama:2007bs}.
By {\sc gencode}{\it N} and its modifications for handling
vacuum-polarization loops and light-by-light-scattering loops,
we have obtained {\sc fortran} programs for 
12 more subsets 
\cite{Aoyama:2008hz,Aoyama:2010zp,Aoyama:2010pk,Aoyama:2011rm,Aoyama:2011zy,Aoyama:2011dy}.
The subsets III(c) and I(j), which involve one(two) light-by-light 
scattering subdiagram(s)  internally,  were handled manually \cite{Aoyama:2008gy,Aoyama:2012fc}. 
The subset II(e), which contain a sixth-order light-by-light-scattering subdiagarm internally, 
was handled by an automation procedure \cite{Aoyama:2010yt}.
At least two independent codes for non-automated programs 
were written by different members of our collaboration 
in order to minimize human errors.

All integrals were numerically evaluated by VEGAS \cite{Lepage:1977sw}.
For some diagrams of  the sets IV and V  that contain
cancellation of linear IR divergence within a diagram, we used  
the quadruple-precision arithmetics  
to avoid possible round-off errors of numerical calculations.
The contribution of the tau-particle loop to $a_e$ is negligible at present.
Thus the sum of (\ref{Eq:A_1(10)}) and (\ref{Eq:A_2(10)}) gives
effectively the total tenth-order QED contribution to $a_e$.

\begingroup
\renewcommand{\baselinestretch}{1.17}
\begin{table}
\caption{
Summary of contributions to the tenth-order lepton $g-2$
from 32 gauge-invariant subsets.
 $n_F$ is the number of vertex 
diagrams contributing to $A_1^{(10)}$.
The numerical  values of individual subsets were 
originally obtained in the references in the fifth column.
The values $A_1^{(10)}$ of  subsets I(d), I(f), II(a), II(b), and VI(c) in
\cite{Kinoshita:2005sm} are corrected as indicated by the asterisk.
The corrected values are listed  in this table.
 \label{Table:all-sets-1}
}
\begin{ruledtabular}
\begin{tabular}{l@{\hskip+0.2em}r@{\hskip-4em}d@{\hskip-6em}d@{\hskip0em}c}
   set  &  $n_F$  & 
\makebox[-5em]{$A_1^{(10)}$} & 
\makebox[-5em]{$A_2^{(10)}(m_e/m_\mu)$ } & 
reference
\\
\hline
 I(a)  &   1 & 0.000~470~94~(6) & 0.000~000~28~(1) &\cite{Kinoshita:2005sm} \\
 I(b)  &   9 & 0.007~010~8~(7)  & 0.000~001~88~(1) &\cite{Kinoshita:2005sm} \\
 I(c)  &   9 & 0.023~468~(2)    & 0.000~002~67~(1) &\cite{Kinoshita:2005sm} \\
 I(d)  &   6 & 0.003~801~7~(5)  & 0.000~005~46~(1) &\cite{Kinoshita:2005sm}$^\ast$ \\
 I(e)  &  30 & 0.010~296~(4)    & 0.000~001~60~(1) &\cite{Kinoshita:2005sm} \\
 I(f)  &   3 & 0.007~568~4~(20) & 0.000~047~54~(1) &\cite{Kinoshita:2005sm}$^\ast$ \\
 I(g)  &   9 & 0.028~569~(6)    & 0.000~024~45~(1) &\cite{Aoyama:2008hz} \\
 I(h)  &  30 & 0.001~696~(13)   &-0.000~010~14~(3) &\cite{Aoyama:2008hz} \\
 I(i)  & 105 & 0.017~47~(11)    & 0.000~001~67~(2) &\cite{Aoyama:2010zp} \\
 I(j)  &   6 & 0.000~397~5~(18) & 0.000~002~41~(6) &\cite{Aoyama:2008gy}\\
 II(a) &  24 &-0.109~495~(23)   &-0.000~737~69~(95)&\cite{Kinoshita:2005sm}$^\ast$ \\
 II(b) & 108 &-0.473~559~(84)   &-0.000~645~62~(95)&\cite{Kinoshita:2005sm}$^\ast$ \\
 II(c) &  36 &-0.116~489~(32)   &-0.000~380~25~(46)&\cite{Aoyama:2011rm} \\
 II(d) & 180 &-0.243~00~(29)    &-0.000~098~17~(41)&\cite{Aoyama:2011rm} \\
 II(e) & 180 &-1.344~9 ~(10)    &-0.000~465~0~(40) &\cite{Aoyama:2010yt} \\
 II(f) &  72 &-2.433~6~(15)     &-0.005~868~(39)   &\cite{Kinoshita:2005sm} \\
 III(a)& 300 & 2.127~33~(17)    & 0.007~511~(11)   &\cite{Aoyama:2011zy} \\
 III(b)& 450 & 3.327~12~(45)    & 0.002~794~(1)    &\cite{Aoyama:2011zy} \\
 III(c)& 390 & 4.921~(11)       & 0.003~70~(36)    &\cite{Aoyama:2012fc} \\
 IV    &2072 &-7.7296~(48)      &-0.011~36~(7)     &\cite{Aoyama:2011dy} \\
 V     &6354 &10.09~(57)        & 0               & Eq.~(\ref{Eq:A_1(10)SetV}) \\ 
 VI(a) & 36  & 1.041~32~(19)    &  0.006~152~(11)  &\cite{Kinoshita:2005sm} \\
 VI(b) & 54  & 1.346~99~(28)    &  0.001~778~9~(35)&\cite{Kinoshita:2005sm} \\
 VI(c) & 144 &-2.5289~(28)      & -0.005~953~(59)  &\cite{Kinoshita:2005sm}$^\ast$ \\
 VI(d) & 492 & 1.8467~(70)      &  0.001~276~(76)  &\cite{Aoyama:2010pk} \\
 VI(e) & 48  &-0.4312~(7)       & -0.000~750~(8)   &\cite{Kinoshita:2005sm} \\
 VI(f) & 180 & 0.7703~(22)      &  0.000~033~(7)   &\cite{Kinoshita:2005sm} \\
 VI(g) & 480 &-1.5904~(63)      & -0.000~497~(29)  &\cite{Aoyama:2010pk} \\
 VI(h) & 630 & 0.1792~(39)      &  0.000~045~(9)   &\cite{Aoyama:2010pk} \\
 VI(i) & 60  &-0.0438~(12)      & -0.000~326~(1)   &\cite{Kinoshita:2005sm} \\
 VI(j) & 54  &-0.2288~(18)      & -0.000~127~(13)  & \cite{Kinoshita:2005sm} \\
 VI(k) & 120 & 0.6802~(38)      &  0.000~015~6~(40)& \cite{Kinoshita:2005sm} 
\end{tabular}
\end{ruledtabular}
\end{table}
\endgroup
\renewcommand{\baselinestretch}{1.00}


\begin{acknowledgments}
This work is supported in part by the JSPS Grant-in-Aid for
Scientific Research  (C)20540261 and (C)23540331.
T. K.'s work is supported in part by the U. S. National Science Foundation
under Grant NSF-PHY-0757868.
T. K. thanks RIKEN for the hospitality extended to him
while a part of this work was carried out.
Numerical calculations are conducted on 
RSCC and RICC supercomputer systems at RIKEN.
\end{acknowledgments}


\bibliography{bwoarX}

\end{document}